\documentclass[11pt,twoside]{article}
\usepackage{asp2014}

\newcommand{\hzav}[1]{\left[#1\right]}
\newcommand{\szav}[1]{\left\{#1\right\}}
\newcommand{\beff}{$B_{\mathrm{eff}}$}
\aspSuppressVolSlug
\resetcounters

\begin{document}

\title{New tools for finding and testing of weak periodical variability}
\author{Zden\v{e}k Mikul\'a\v{s}ek,$^{1,2}$ Ernst Paunzen,$^1$ Martin Netopil$^1$ and Miloslav Zejda$^1$}
\affil{$^1$Department of Theoretical Physics and Astrophysics, Masaryk University, Brno, Czech Republic; \email{mikulas@physics.muni.cz}}
\affil{$^2$Observatory and Planetarium, V\v{S}B, Technical University, Ostrava, Czech Republic}

% This section is for ADS Processing.  There must be one line per author.
\paperauthor{Sample~Author1}{Author1Email@email.edu}{ORCID_Or_Blank}{Author1 Institution}{Author1 Department}{City}{State/Province}{Postal Code}{Country}
\paperauthor{Sample~Author2}{Author2Email@email.edu}{ORCID_Or_Blank}{Author2 Institution}{Author2 Department}{City}{State/Province}{Postal Code}{Country}
\paperauthor{Sample~Author3}{Author3Email@email.edu}{ORCID_Or_Blank}{Author3 Institution}{Author3 Department}{City}{State/Province}{Postal Code}{Country}

\begin{abstract}
Our paper presents new methods for finding and testing of weak periodic variability of stellar objects developed for the purpose of detecting expected regular light variations of magnetic chemically peculiar (mCP) candidates in the Large Magellanic Cloud. We introduce two new periodograms of the mCP star, BS Cir (HD~125630), appropriate for rotating spotted variables and compare the results with those obtained by the well-known Lomb-Scargle periodogram. The usage of periodograms and the testing of the significance of the found period candidates are demonstrated with two examples: the observed and simulated observations of the magnetic field of the mCP star CQ UMa (HD 119213) and the mCP candidate OGLE LMC136.7 16501. Three newly developed tests of the periodic variability -- the shuffling, bootstrap and subsidiary ones, are presented. We demonstrate that the found periodic variations known with Signal-to-Noise ratio larger than 6 can be approved as real.
\end{abstract}

\section{Introduction}
The chemical peculiar (CP) stars of the upper main sequence display abundances that deviate significantly from the standard (solar) abundance distribution. The existence of strong global magnetic field specifies a subset of this class, the magnetic chemically peculiar (mCP) stars.

The periodic variability of mCP stars is explained in terms of the oblique rotator model, according to which, the period of the
observed light, spectrum, and magnetic field variations is identical to the rotational period. The photometric changes are due to
variations of global flux redistribution caused by the phase-dependent line blanketing and continuum opacity namely in the ultraviolet part of stellar spectra \citep{krt901,krtcuvir}.
The amplitude of the photometric variability is determined by the degree of non-uniformity of the surface brightness (spots), the used passband, and the line of sight. The observed light amplitudes are up to a few tenths of magnitudes, standardly 

In the Milky Way, we know of a statistically significant number of rotational periods for mCP stars deduced from photometric and/or spectroscopic variability studies \citep{reca01,zoo}. Nevertheless, also extragalactic mCP stars were found in the meanwhile.

After the first photometric detection of classical chemically peculiar (CP) stars in the Large Magellanic Cloud (LMC) \citep{mai01}, a long term effort was spent to increase the sample \citep{pau06}. Finally, we were able to verify our findings with spectroscopic observations \citep{sipau}.

In this paper, we present the tools of the time series analysis developed for finding and testing of the expected periodic light variations of mCP candidates in the LMC \citep{pau}. The list of targets \citep{pau06} was compared with the OGLE database \citep{udal} for corresponding measurements. In total, fourteen common objects were found and the $V$ and $I$ light curves analysed. The description of methods are also published in the textbook by \citet{mikzej}.

\section{The periodograms}

The basic tool to search for periodic variations of a measured value (intensity, magnitude, and radial velocity) are so called periodograms. These plot some quantities expressing the significance of the phase sorting of searched data according to an individual angular (circular) frequency $\omega=2\,\pi\,f=2\,\pi/P$, where $f$ is a frequency, $f=P^{-1}$, $P$ is a period. The extremes of periodograms then indicate the best arrangement of observed data versus a period and hence the possible periodicity of a signal.

The basic linear regression models of periodic variations are harmonic polynomials of the $g$-order:
\begin{equation}\label{harm}
F(t,\omega)=\sum_{j=1}^g\,\beta_{1\!j}(\omega)\cos(\omega j\,t)+\beta_{2\!j}(\omega)\sin(\omega j\,t),
\end{equation}
where $F(t,\omega)$ is the model of detrended measured quantities $\szav{y_i(t_i)}$ corrected for their mean, $\beta_{1\!j}(\omega),\,\beta_{2\!j}(\omega)$ are $2\,g$ harmonic coefficients. The harmonic coefficients for the best fit of model function $F(t,\omega)$: $b_{1\!j}(\omega),\,b_{2\!j}(\omega)$ for the fixed $\omega$ can be determined by the standard least square method technique allowing to count with uneven uncertainties of individual measurements $\szav{\sigma_i}$.

\subsection{Periodograms with modulated amplitude}\label{modulator}
The simplest way how to construct LSM spectral periodogram is to plot scalar value $\chi^2(\omega)$ versus $\omega$ or $f=P^{-1}$, where
\begin{equation}\label{chikva}
\chi^2(\omega)=\sum_{i=1}^n\,\hzav{\frac{y_i-F(\omega,t_i)}{\sigma_i}}^2=
\sum_{i=1}^n\, \hzav{\frac{y_i^2}{\sigma_i^2}- \frac{F^2(\omega,t_i)}{\sigma_i^2}}.
\end{equation}
Now we can find and discuss the frequencies for which the value $\chi^2(\omega)$ reach their minima. This method is fairly general because it can be applied to any kind of time series (magnitudes, intensities, spectral line equivalent widths, or radial velocities). Nevertheless for data of the same type (magnitudes, intensities) we recommend to use the following modification with some 'value added'.

The first sum of equation (\ref{chikva}) where the first sum on the right is a constant that not depends on the frequency, while the second is the weighted sum of the squares of the model prediction given by the function $F(\omega,t)$. Therefore, instead of the minimum of the $\chi^2(\omega)$ we can find the maximum of the modulated amplitude $A_{\mathrm{m}}$
\begin{equation}
A_{\mathrm{m}}^2=\frac{8}{\sum_{i=1}^n\sigma_i^{-2}}\sum_{i=1}^n \, \frac{F^2(\omega,t_i)}{\sigma_i^2},\label{Amgen}
\end{equation}
which is nearly equal to the effective amplitude $A_{\mathrm{eff}}(\Phi)$ of a periodic function \citep[see in][]{mikan}.

For the first estimate of the variation period it is sufficient to use the simplest possible option: $g=1$ which gives also the most explicit results. Then
\begin{equation}\label{modul}
A_{\mathrm{m}}^2(\omega)=\frac{8}{\sum_{j=1}^{n} \sigma_j^{-2}} \displaystyle \sum_{i=1}^n \hzav{\frac{b_{11}(\omega)\cos(\omega\,t_i)+ b_{21}(\omega)\sin(\omega\,t_i)}{\sigma_i}}^2.
\end{equation}
\begin{figure}[h]
\centerline{\includegraphics[width=0.98\textwidth]{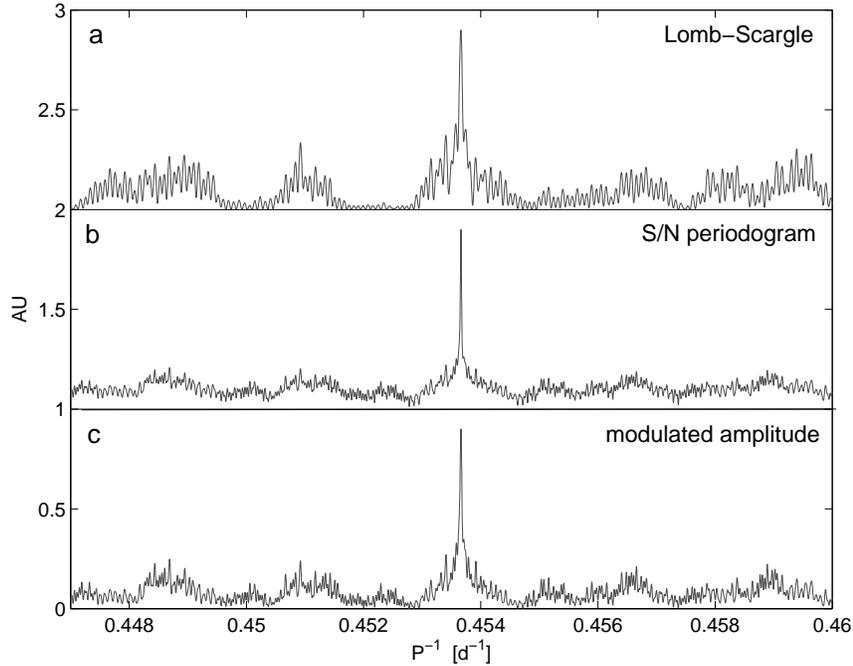}}
\begin{center}
\caption{\small All three periodograms of the moderately cool chemically peculiar star BS Cir \citep[HD 124224; for more information see e.g. in][]{mik13,mik15} undoubtedly pinpoint the only dominant period peak at $P=2\fd2042$. Periodograms are displayed in arbitrary units normalized to the height of the period peak. The results of particular periodograms are comparable, nevertheless both of the new tools of periodograms ((b) and (c); Sec.\,\ref{sumator} and Sec.\,\ref{modulator}) are a slightly better than the results of the notorious Lomb-Scargle periodogram ((a), Sec.\,\ref{Scargle}).} \label{Fig1}
\end{center}
\end{figure}

\begin{figure}[h]
\centerline{\includegraphics[width=0.97\textwidth]{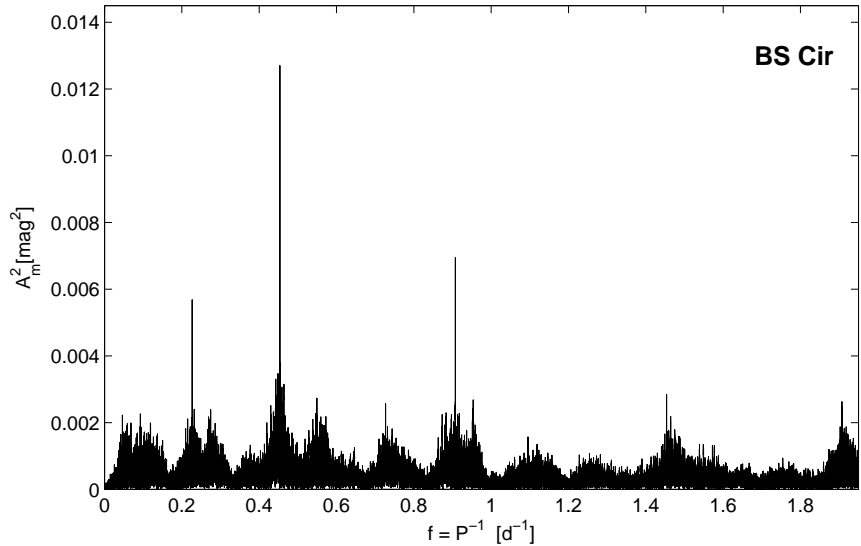}}
\begin{center}
\caption{\small The periodogram of the square of the modulated amplitude, constructed on the basis of 4766 individual photometric observations in the spectral interval 505 to 560 nm, indicates the main frequency $f_0=P^{-1}=0.4538$\,d$^{-1}$ (rotation period) and subsidiary peaks at  $f_0/2,\,2f_0,\,3f_0,\,4f_0.$} \label{Fig2}
\end{center}
\end{figure}

\subsection{Robust S/N periodogram}\label{sumator}

The second LSM type of periodograms uses as a criterion of the significance of individual peaks a
robust ``signal-to-noise'' $S/N$ criterion which is defined as:
\begin{equation}\label{svist}
\displaystyle S/N(\omega)=\frac{Q(\omega)}{\delta Q(\omega)},\quad \mathrm{where}\quad Q=\sum_{j=1}^g b_{1j}^2+b_{2j}^2.
\end{equation}
$\delta Q(\omega)$ is an estimate of the uncertainty of the quantity $Q(\omega)$ for a particular angular frequency. The model function $F(\omega,t,b_{11},b_{21},\ldots)$ is assumed in the form of the harmonic polynomial of the $g$-th order (see Eq.\,\ref{harm}). The detailed description of both LSM novel periodogram criteria can be found in the textbook by \citet{mikzej}.

We tested the properties of the $S/N(\omega)$ criterion on thousands samples with sine $(g=1)$ signals scattered by randomly
distributed noise. We found that if there is no periodic signal in such data, the median of the maximum $S/N(\omega)$ value
in a periodogram is 4.52, in 95\% of cases we find a $S/N$ value between 4.2 and 5.4. Consequently, the occurrence of peaks definitely higher than 6 indicates possible periodic variations.

The periodogram $S/N(\omega)$ is very sensitive (see Fig.\ref{Fig1}b) and suppresses the false periods which results from usually bad time distribution of astrophysical observational data, very well.

\subsection{The classical Lomb-Scargle periodogram}\label{Scargle}

During the treatment of OGLE-III time series \citep{pau}, we concluded that both types of periodograms correlate very well with other time-proven periodograms as e.g. the Lomb-Scargle \citep[see e.g.][]{rybicki} one. So we are able to consider them as generally interchangeable (see also Fig.\,\ref{Fig1}).

The Lomb-Scargle method assumes that the found changes have sine/cosine type phase curves. The method uses, as the measure of individual periods significance, the quantity $Q_{\mathrm{LS}}(\omega)$:
\begin{eqnarray}\label{lomb}
\displaystyle &\tau=\omega^{-1}\arctan\hzav{ \displaystyle\frac{ \sum_{i=1}^n\sin(2\,\omega\,t_i)} {\sum_{i=1}^n\cos(2\,\omega\,t_i)}}; \quad \displaystyle \omega=\frac{2\,\pi}{P};\\
& Q_{\mathrm{LS}}(\omega)=\displaystyle \frac{ \sum_{i=1}^n \szav{y_i\, \cos[\omega\,(t_i-\tau)]}^2}{\sum_{i=1}^n \cos^2[\omega\,(t_i-\tau)]}+\displaystyle \frac{\sum_{i=1}^n \szav{y_i \sin[\omega\,(t_i-\tau)]}^2}{\sum_{i=1}^n \sin^2[\omega\,(t_i-\tau)]}.\nonumber
\end{eqnarray}
The definition of the significance indicator $Q_{\mathrm{LS}}$ (see Eq.\,\ref{lomb}) can be easily modified to the more general case of data with uneven uncertainties.

\section{Examples of strong and weak variability}

\subsection{Light variation of BS Cir - a well observed object with strong light variations}

BS Cir (HD 125630) is a moderately cool mCP star with the rotation period $P=2\fd24$.  There are many observations of the double-wave light curves available \citep[for details see in][]{mik13,mik15}. Although light curves taken in various colours differ, those ones obtained in the spectral region 500--560 nm can be considered as almost identical.

The periodogram\footnote{The periodograms of BS Cir were based on 4766 individual \textit{Hp, y} and $V$ measurements taken from \citet{vofa79,mare83,cale93,esa,pojm,omc,mik13,mik15}.} presented in Fig.\,\ref{Fig2} shows the spectrum of the square of the modulated amplitude $A_{\mathrm{m}}^2(f)$, $(g=2$, see Eqs.\,\ref{harm},\,\ref{Amgen}) dominated by the basic period with frequency $f_0=0.45363$\,d$^{-1}$, and subordinate peaks at  $f_0/2,\,2f_0,\,3f_0,4f_0$. This spectrum results from the fact that the light curves in various filters are not exactly the same. The mean modulated amplitude of the maximum peak correspond to 0.11 mag, the mean weighted $S/N$ of used 4766 individual photometric observations is about 20.

\begin{figure}[h]
\begin{center}
\includegraphics[width=0.8\textwidth]{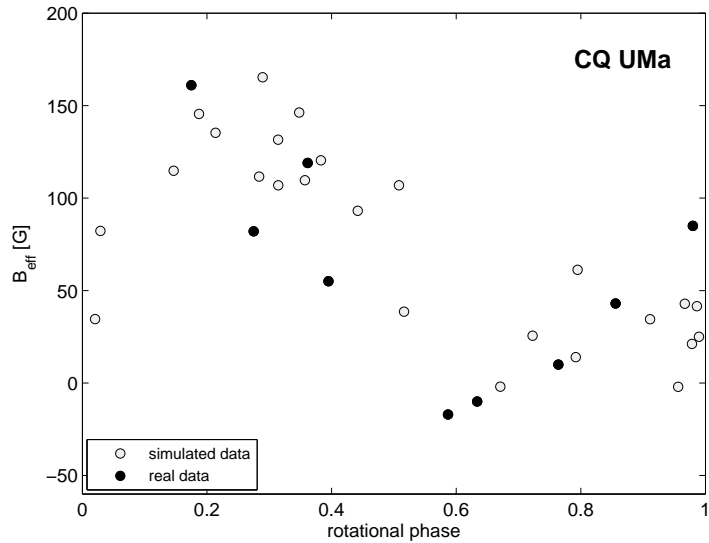}
\caption{\small The phase diagram of real and artificial (simulated) magnetic field observations CQ UMa. The periodogram of based on 9 measurements of its magnetic field obtained during 2 years. Real rotational period is denoted by the arrow.} \label{Fig3}
\end{center}
\end{figure}

We obtain such a periodogram only in very rare cases when we study an object with the pronounced variability documented by thousands of observations not suffering from bad sampling. Nevertheless, frequently we have to analyze data of much worse quality and time sampling.

\subsection{Magnetic field of CQ UMa - strong variability, but few observations}

CQ UMa (HR\,5153 = HD 119213), is also a moderately cool mCP star, similar to BS Cir, with a strong variation in the $v$ color and $c_1$ index. The period of variations of all kind is constant, $P = 2\fd449967$ \citet{unsteady}.
\begin{figure}[h]
\begin{center}
\includegraphics[width=0.8\textwidth]{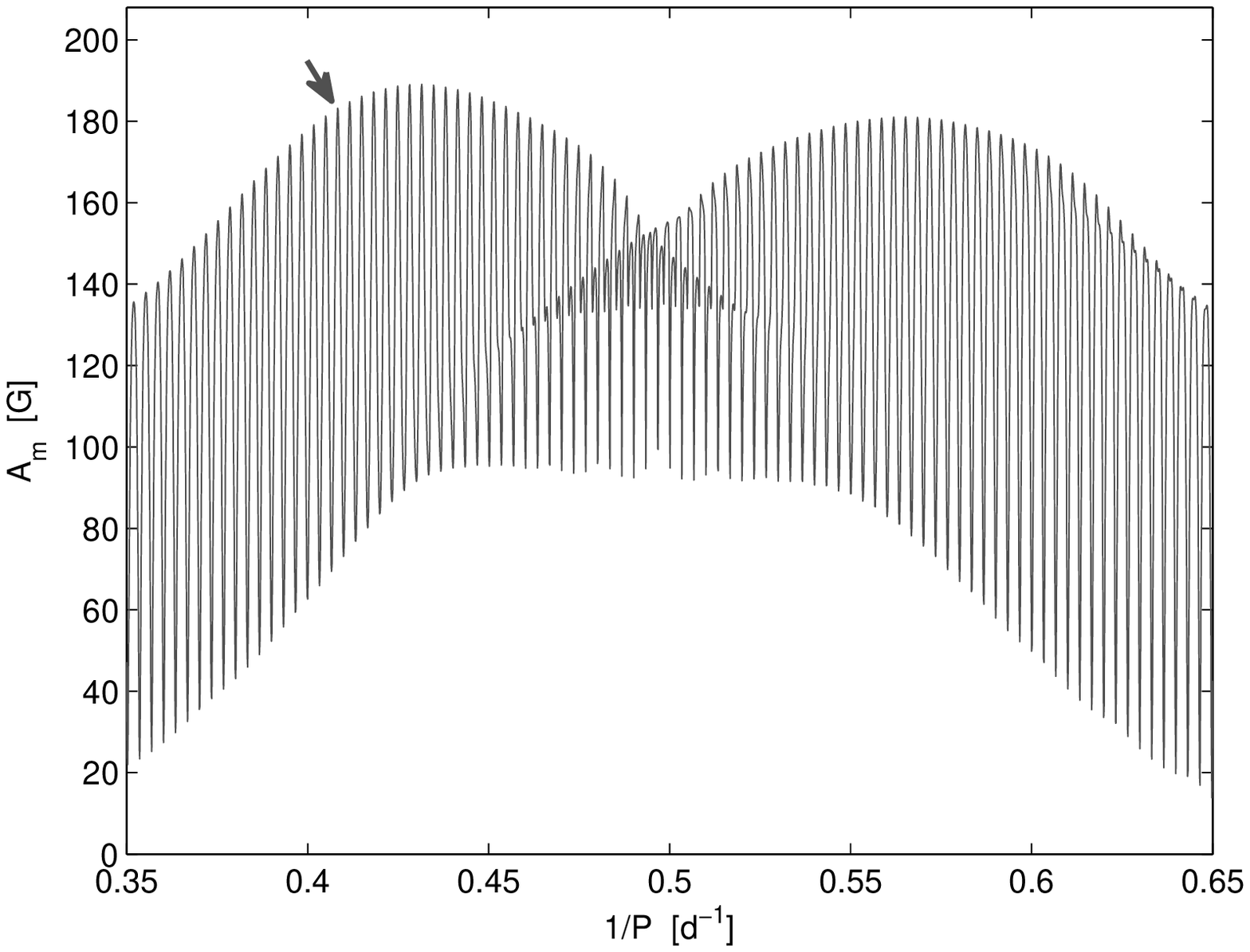}
\caption{\small The periodogram of CQ UMa based on 9 measurements of its magnetic field obtained during 2 years. Real rotational period is denoted by the arrow.} \label{Fig4}
\end{center}
\end{figure}
\begin{figure}[h]
\centerline{\includegraphics[width=0.8\textwidth]{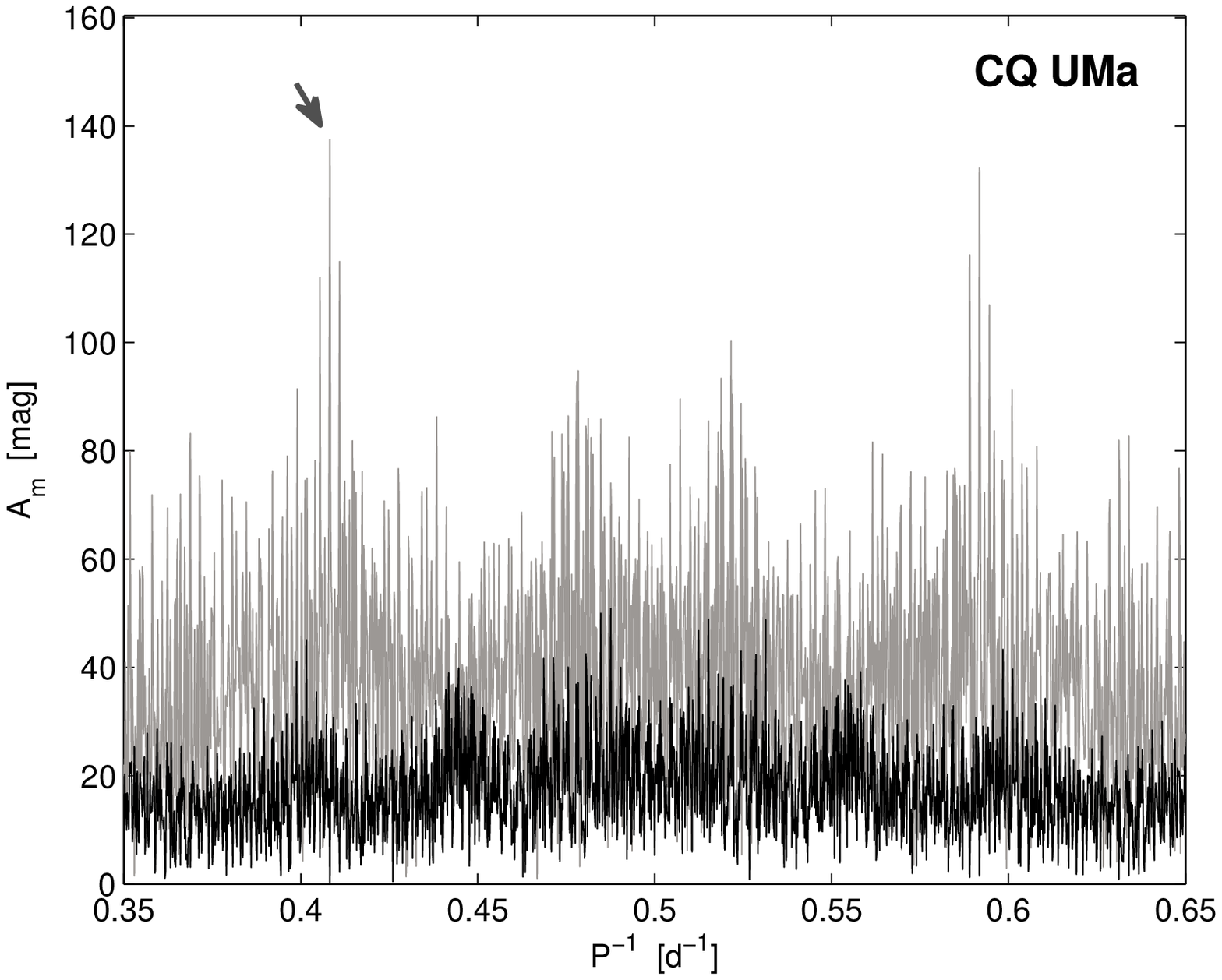}}
\begin{center}
\caption{\small The periodogram of CQ UMa based on 27 measurements of its magnetic field obtained during 6 years. Real rotational period is denoted by the arrow. The dark periodogram corresponds to the period analysis if we subtract the variations with the basic period $P=2\fd449967$. The low amplitude of it and the absence of any other pronounced period peaks mean that the periodogram of the star is formed namely by aliases of the basic (true) period. } \label{Fig5}
\end{center}
\end{figure}

In 1983-4, we obtained nine measurements of the effective magnetic field ($B_{\mathrm{eff}}$) using the hydrogen magnetograph of the 6-m SAO reflector (see full dots in Fig.\,\ref{Fig3}). The expected synchronous periodic variations in the light and magnetic field ($B_{\mathrm{eff}}$) were confirmed \citep{mik84}.

However, are these nine magnetic observations of CQ UMa sufficient for the period determination? Definitely not, the period spectrum is overcrowded by plenty of aliases (Fig.\,\ref{Fig4}). Several of them are so much dominant that they exceed even the peak of the real period.

The aliases can be suppressed by a continuation observations. We can simulate it by adding of 27 `new' (artificial) magnetic measurements obtained during the following six years (see open circles in Fig.\,\ref{Fig3}). The new periodogram (Fig.\,\ref{Fig5}) undoubtedly indicates the only dominant period at the true placement\footnote{The other dominant peaks in the periodogram are aliases conjugated with the period of one sidereal day through the Tanner relation -- see Eq.\,\ref{Tanner}}. The phase diagram plotted with this period displays simple sinusoidal variations with a $S/N\sim 8$.

If we remove the sinusoidal signal from our magnetic measurements, we can search for other possible relic periods using the same periodogram technique - see the bottom of Fig.\,\ref{Fig4}. It seems that there is no periodic signal in the remaining data meaning that the found period is unique. The remaining data correspond to pure noise which is not very prominent. The characteristics of the periodogram is determined mainly by aliases originated from the very low number of measurements (36) and the short time basis of observations (8 years).

The aliases can be lowered if we use, besides not very accurate magnetic measurements of \beff, more precise photometric data ($S/N$ usually up to 20). Then we are able to achieve the pure periodogram similar to the periodogram of BS Cir displayed in Fig.\,\ref{Fig2}.

\subsection{Is the mCP candidate OGLE LMC136.7 16501 periodically variable or not?}

Several quite different problems were encountered when we analyzed the light curves of mCP candidates in the Large Magellanic Cloud. The number of individual OGLE observations in the $V$ and $I$ colours are $\sim 300-400$ and the duration of observations (8 years) were satisfactory. The effect of aliases in such data should be insignificant. Unfortunately, the periodograms suffer here from relatively large scatter due to observational noise. The ratio $S/N$ of an individual observation was very low because of the weakness of the signal.

We shall now discuss the period analysis of a typical representative of the mCP candidates in LMC - the star OGLE LMC136.7 16501 (04 50 46.10 -69 59 16.7, 2000.0), denoted in \citet{pau} as a star No. 4, with a mean magnitude in $V = 19.16$ mag, the absolute magnitude $M_V = 0.41$ mag.
\begin{figure}[h]
\centerline{\includegraphics[width=0.90\textwidth]{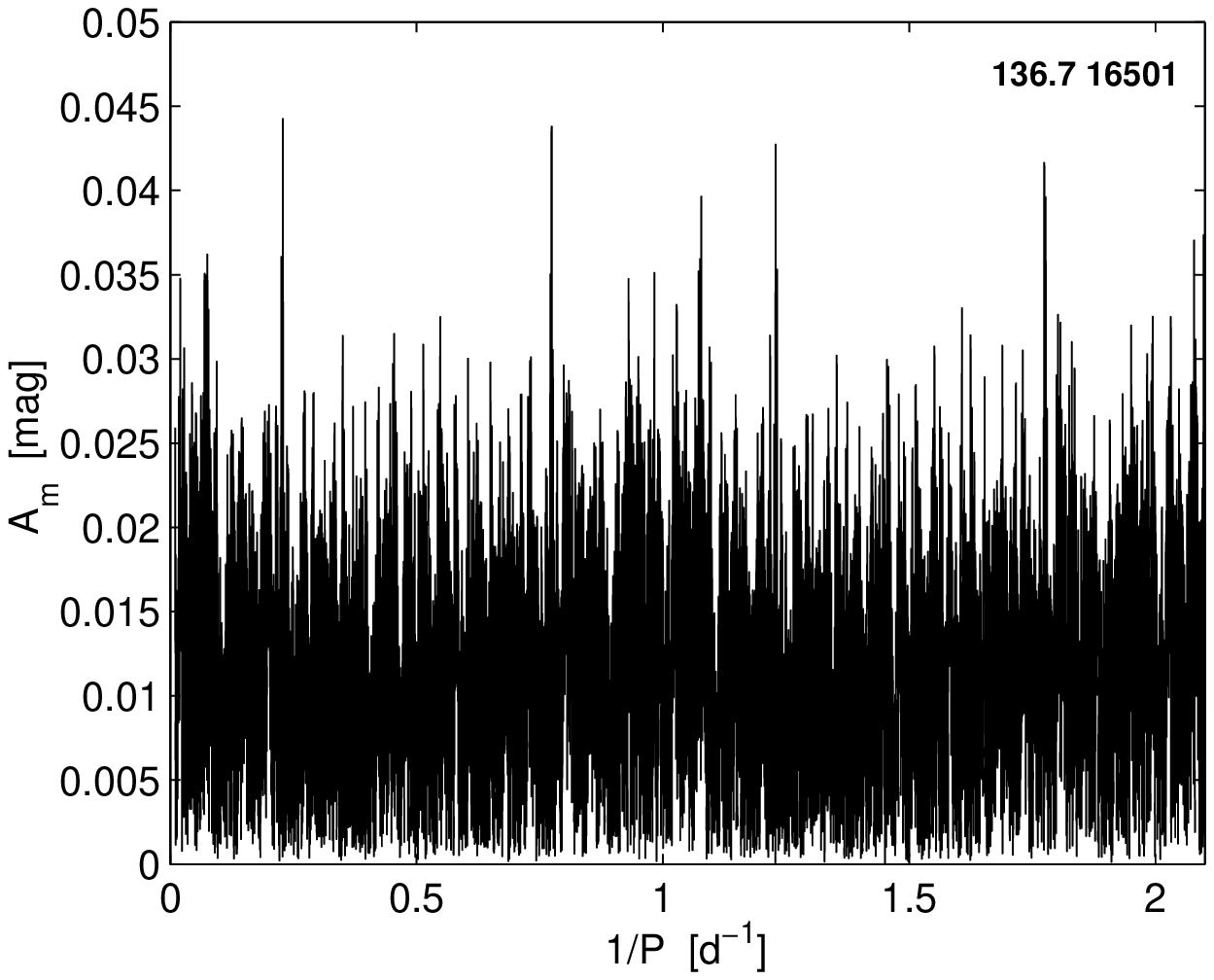}}
\begin{center}
\caption{\small The periodogram of LMC 136.7 16501 displays many period peaks. The period spectrum corresponds to nearly pure scatter.} \label{Fig6}
\end{center}
\end{figure}
\begin{figure}[h]
\centerline{\includegraphics[width=1\textwidth]{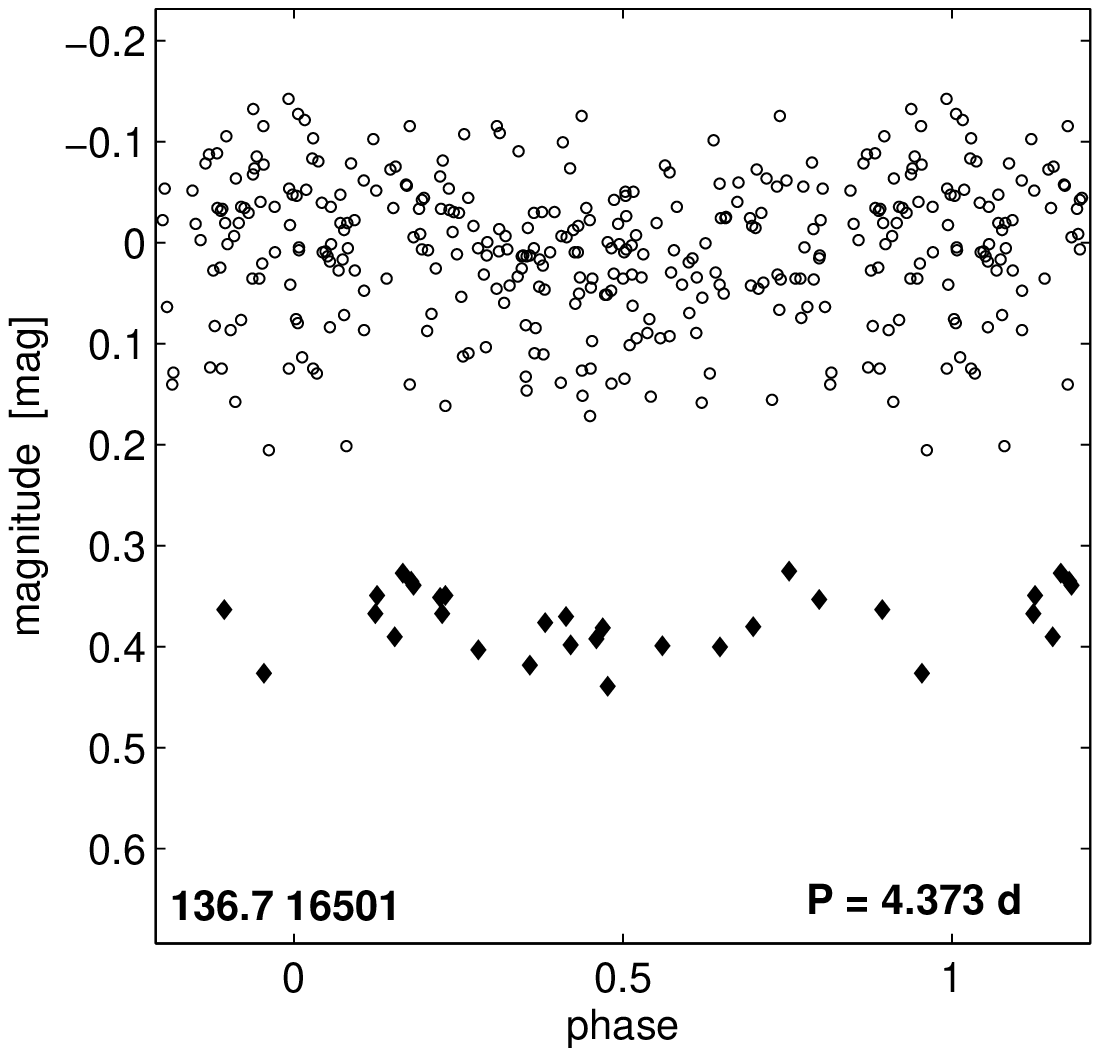}}
\begin{center}
\caption{\small $I$ and $V$ light curves of OGLE LMC136.7 16501 plotted versus the `best' period $P=4\fd373$.} \label{Fig7}
\end{center}
\end{figure}
The periodogram depicted in Fig.\,\ref{Fig6} is quite dissimilar to the periodograms of mCP stars discussed above (compare with Fig.\,\ref{Fig2} and Fig.\,\ref{Fig5}). We do not see here any  prominent period peak. The characteristics of the period spectrum is mainly determined by the stochastic scatter that competes with several possible periods of the the principal aliases conjugated with the basic sampling period of one sidereal day. The frequencies of aliases obey to notorious \citet{tan} relation:
\begin{equation}\label{Tanner}
f_{\mathrm{al}}(k)=|f_{\mathrm{M}}+k\,f_{\mathrm{s}}\,|,\quad \mathrm{where}\quad k=(\ldots -2,-1,1,2,\ldots),
\end{equation}
where $f_{\mathrm{al}}(k)$ is the frequency of the alias of the $k$-th order, $f_{\mathrm{M}}$ is the frequency of the maximum period peak $(f_{\mathrm{M}}=f_{\mathrm{al}}(0))$, and $f_{\mathrm{s}}$ is the sampling frequency, here it is the frequency of a sidereal day $f_{\mathrm{s}}=1.00274\,\mathrm{d}^{-1}$.

The most significant period in the periodogram (Fig.\,\ref{Fig6}) is 4\fd373, with amplitudes of the light curves $A_{\mathrm m} = 0.044$ mag, both in $V$ and $I$. The ratio of $S/N = 5.1$ for this peak is very low. The light curves in $V$ and $I$ are depicted in Fig.\,\ref{Fig7}. Nevertheless, all predicted aliases: $f_{\mathrm{al}}(-2)=1.777,\,f_{\mathrm{al}}(-1)=0.774,\, f_{\mathrm{\mathrm{al}}}(1)=1.231,\,f_{\mathrm{al}}(2)=2.234$ are indicated.

Does it mean that the period 4\fd373 is real? Unfortunately not. The occurrence of aliases at their `right positions' given by the Tanner relation Eq.\,\ref{Tanner} is only the confirmation of the distinctiveness of expected sampling period or periods in observational data. Because the OGLE data were obtained at a ground based station, the pronounced one day sampling period can be easily explained. The observed ordering of observation into a phase curve with the period 4\fd373 can be a mere coincidence.

That is why we shall solve the crucial question: Are the found periodic variations real? Some partial answers may give us several simple tests of variability of general use we have developed \citep{pau}.

\subsubsection{Shuffling method}\label{shuple}

The heuristic \emph{shuffling method} is able to test the `pessimistic' hypothesis that the distribution of the data is only random. We can analyze the data of our object displaying weak periodical variations again by the same way as it was described above, only we randomly shuffled all the individual observations
(magnitudes and their uncertainties), the times of the observations remained the same. We are convinced that this derogation will destroy any periodic signal and randomize the data. For the shuffled version of data we then find the maximum modulated amplitude $A_{\rm{m}}$ and the maximum of $S/N$. Then we repeat the same procedure many times and compare the results with $A_{\rm{m}}$ and $S/N$ for the original unshuffled data set. If the results are nearly the same, we may conclude that the periodic variability of the object (if any) is undetectable in the investigated data.

We found for the mCP candidate OGLE LMC136.7 16501, the median of amplitudes in period peaks of shuffled data $\tilde{A}_{\mathrm m} = 0.040$ mag and a ratio $\widetilde{S/N} = 4.6$, what is very close to the values found above $A_{\mathrm m} = 0.044$ mag, both in $V$ and $I$ and $S/N = 5.1$. Consequently, the LMC136.7 16501 photometric data are very probably random.

The same results we obtained also for the other 11 mCP candidates, only for stars No 12 and No 14 \citep[according to the list of LMC mCP cadidates in][]{pau} the periods seem significant. We deduce a $S/N > 6$ for significance.

The following tests will help to quantify the significance of the periods.

\subsubsection{Bootstrap test}

The technique of bootstrap \citep{hall92} has proved to be very useful for testing the statistical significance and
therefore the reality of found periods. It helped us to quantify this reality as a probability that the periodogram of a randomly
created bootstrap subset has its dominant peak at the same frequency as the standard periodogram. We tested it with one hundred of
bootstrap subsets for each star of our sample. We consider a period as statistically significant, if the maximum peak occurs at one
of the aliased frequency because during the bootstrap choice aliases often exceed the basic peak.

The results for found periods of all 14 mCP candidates were dismal: the bootstrap significance never exceeds 50\%. Especially, the bootstrap significance of the period 4\fd37 in the case of the star 4 was only 26\%! The periodicity of stars No. 12 and 14, mentioned in Sec.\,\ref{shuple}, is disputable.

\subsubsection{Subsidiary test}
For data, including observations in two or more spectral regions -- their individual periodograms should indicate nearly the same peak period. It can be tested, e.g. by the periodogram of the product of individual $A_{\mathrm{m}}$ or $S/N$ values.
\section{Instead of a conclusion}
All mentioned newly developed methods are open source. For details see in \citet{pau} or the textbook of \citet{mikzej}.

\acknowledgements This paper uses observations made at the South African Astronomical Observatory (SAAO).
This work was funded by the grant GA\v{C}R P209/12/0217 and
supported by the SoMoPro II Programme (3SGA5916),
co-financed by the European Union and the South Moravian Region, the
grant GA \v{C}R 7AMB12AT003, and
the financial contributions of the Austrian Agency for International
Cooperation in Education and Research (BG-03/2013 and CZ-09/2014). 
MN acknowledges the support by the grant 14-26115P of the Czech Science Foundation.This work reflects the opinion of the authors and the European Union is not responsible for any possible application of the information included in the paper.

\end{document}